\documentclass[11pt]{article}
\usepackage{amssymb}

\def\dd{\mbox{d}}

\def\O{\Omega}
\def\o{\omega}

\def\a{\alpha}
\def\b{\beta}
\def\d{\delta}
\def\D{\Delta}

\def\g{\gamma}

\def\e{\epsilon}

\def\f{\phi}
\def\F{\Phi}

\def\l{\lambda}
\def\L{\Lambda}
\def\m{\mu}
\def\n{\nu}
\def\s{\sigma}
\def\S{\Sigma}
\def\o{\omega}

\def\r{\rho}
\def\t{\tau}
\def\th{\theta}

\def\pa{\partial}

\newcommand{\ti}[1]{\tilde{#1}}

\newcommand{\sm}[1]{\mbox{\scriptsize #1}}
\newcommand{\tn}[1]{\mbox{\tiny #1}}
\renewcommand{\@}[1]{\sqrt{#1}}
\newcommand{\Tr}{{\mbox{Tr}}\,}
\renewcommand{\le}[1]{\label{#1}\end{eqnarray}}
\newcommand{\be}{\begin{equation}}
\newcommand{\ee}{\end{equation}}
\newcommand{\bea}{\begin{eqnarray}}
\newcommand{\eea}{\end{eqnarray}}
\newcommand{\nn}{\nonumber}

\newcommand{\eq}[1]{(\ref{#1})}
\def\nn{\nonumber\\}

\def\ffract#1#2{\raise .35 em\hbox{$\scriptstyle#1$}\kern-.25em/
\kern-.2em\lower .22 em \hbox{$\scriptstyle#2$}}

\def\half{{1\over2}\,}
\begin{document}

\begin{flushright}
AEI-2004-056\\
{\tt hep-th/0407139}\\
\end{flushright}
\vskip0.1truecm

\begin{center}
\vskip 2.5truecm {\Large \textbf{Chern-Simons Theory in Lens Spaces from\\
$ $\\
2d Yang-Mills on the Cylinder}}
\vskip 1.5truecm

{\large\bf{Sebastian de Haro}}\\
\vskip .9truecm {\it Max-Planck-Institut f\"{u}r
Gravitationsphysik\\
Albert-Einstein-Institut\\
14476 Golm, Germany}\\
\tt{sdh@aei.mpg.de}

\end{center}

\vskip 2truecm

\begin{center}
\textbf{\large Abstract}
\end{center}

We use the relation between 2d Yang-Mills and Brownian motion to show that 
2d Yang-Mills on the cylinder is related to Chern-Simons theory in a class of
lens spaces. Alternatively, this can be regarded as 2dYM computing certain 
correlators in conformal field theory. We find that the partition function of 
2dYM reduces to an operator of the type $U=ST^pS$ in Chern-Simons theory for 
specific values of the YM coupling but finite $k$ and $N$. $U$ is the operator 
from which one obtains the partition function of Chern-Simons on 
$S^3/{\mathbb{Z}}_p$, as well as expectation values of Wilson loops. The 
correspondence involves the imaginary part of the Yang-Mills coupling being
a rational number and can be seen as a generalization of the relation between
Chern-Simons/WZW theories and topological 2dYM of Witten, and Blau and 
Thompson. The present reformulation makes a number of properties of 2dYM on 
the cylinder completely explicit. In particular, we show that the modular 
transformation properties of the partition function are intimately connected 
with those of affine characters.

\newpage

\section{Introduction and summary of results}

Two-dimensional Yang-Mills is a well-studied theory \cite{migdal,gross,moore}, 
especially in its
large $N$ limit where it can be reformulated as a string theory \cite{gross}. 
Being a nice playground
to study quantum gauge theory, both perturbatively and non-perturbatively, as well as 
being able to explicitly manifest itself as a string theory, this theory very explicitly 
realizes the 
idea of 't Hooft \cite{thooft} that gauge theories at large $N$ can be reformulated as  
string theories, and has been an object of extensive study for the last 
twenty years or so. To the extent that one might think there are no interesting questions for Yang-Mills left 
to answer on a Riemann surface.

Yet there are several indications that much physics is still to be extracted from bosonic
2dYM. Let us mention just a few of them. In \cite{vafa}, it was shown that 2d Yang-Mills 
on the torus is related to a string theory at large $N$, but of a different kind from the
one
found in \cite{gross,grosstaylor}. It is not defined on a two-dimensional 
target space, but rather as the topological A-model on a six-dimensional non-compact Calabi-Yau. The correspondence was argued both on the topological vertex side and on the
4d SYM theory side, where the twisted 4d SYM reduces
to 2d bosonic YM. Related ideas had been discussed previously in \cite{robbert}.
Vafa has further conjectured \cite{vafa} that 2dYM may have some relevance in
the description of 4d black holes of type II superstrings.

Nekrasov \cite{nekrasov} has shown that random partitions can be used to compute instanton 
contributions to the prepotential of ${\cal N}=2$ SYM theory. In \cite{brownian}, it was shown
that these partitions are related to, among other things, Wilson's plaquette model of QCD$_2$ at
finite $N$. In \cite{mmo}, random partitions were shown to be related to large $N$ 2dYM on the sphere.

In 2dYM, the partition function of the sphere and the torus can be obtained from 
that of the cylinder by sending the external states to zero and by tracing over them,
respectively. Therefore, the above suggests that a reexamination of 2dYM on the cylinder 
might be desirable.

Finally, in \cite{brownian} a relation was found between Brownian motion in a Weyl chamber and Chern-Simons theory, and between Brownian motion in a Weyl alcove and 2d Yang-Mills. This suggested some type of relation between Chern-Simons and 2d Yang-Mills. Such relations had already been pointed out in \cite{witten2d,witten,dv}.

In this paper we reformulate 2d Yang-Mills on the cylinder along the lines of 
\cite{brownian}. We  use the techniques of earlier studies of the heat equation in the mathematical literature \cite{frenkel,fegan1}. This makes the modular properties of the theory completely explicit. In \cite{douglas,rudd}, these modular properties were studied for the torus. The chiral theory was found to transform with weight $6g-6$ in the 't Hooft expansion, but the non-chiral theory did not seem to have definite transformation properties. On the cylinder, we can reformulate the partition function of 2d Yang-Mills in terms of 
generalized theta-functions and work out its modular properties explicitly.

The new formulation of 2d Yang-Mills also sheds some light on the relation between 2dYM
and Chern-Simons. We find that for special imaginary values of the 
coupling ($2\pi$ times a rational number) and particular external states, corresponding 
to integrable representations, the partition function equals the gluing operator 
$U=ST^pS$ in Chern-Simons theory, from which the partition function on 
$S^3/{\mathbb{Z}}_p$ and expectation values of up to two Wilson loops can be 
obtained. Notice that the partition function of 2d Yang-Mills diverges at these
particular values of the coupling. This divergence is analogous to those found in
\cite{witten,blauthompson} at zero coupling, and reflects the fact that 2dYM is
generically non-analytic in the coupling.

In another line of developments, Witten \cite{witten2d,witten} and Blau and Thompson
\cite{blauthompson} showed that topological 2dYM/BF theory is related to Chern-Simons
theory on certain spaces, and in particular computes the dimension of the space of
conformal blocks in the WZW model. Topological 2dYM is simply the $g_{\tn{YM}}\rightarrow
0$ limit of 2dYM, where the partition function reduces to the volume of the moduli space
of flat connections on a Riemann surface. It is equivalent to BF theory with vanishing
potential. This can be compared to CFT only in the limit
$k\rightarrow\infty$. To go beyond the $k\rightarrow\infty$ limit, Blau and Thompson
\cite{blauthompson} showed that one can use the same topological BF theory but with a 
compactified scalar. The limit $k\rightarrow\infty$ then corresponds to the limit where 
the scalar decompactifies. This theory however did not appear to have a straightforward
Yang-Mills interpretation for finite $k$.

Our results could be interpreted as a 2d Yang-Mills description of the finite $k$ case,
giving some sort of topological sector of 2dYM for special values of the coupling.
Our theory can be reformulated as an ordinary BF theory (i.e., with non-compact scalar)
with a quadratic potential and complex coupling.

The organization of the paper is as follows. In section 2 we analyze the structure of 
2dYM on the cylinder by reformulating it in terms of Brownian motion and obtain the 
expressions that are used in the remainder of the paper. A reader who is 
mainly interested in the relation to Chern-Simons/WZW model might want to have only a 
quick glance at this section. In section 3 we show that complexifying
the Yang-Mills coupling the partition function of 2dYM
is equal to the (appropriately normalized) operator
$U=ST^qS$. In section 4 we discuss how our results relate to those by Witten, and Blau 
and Thompson where they compute dimensions of moduli spaces of flat connections by using 
Chern-Simons theory. We end with some conclusions and open problems.

\section{Two-dimensional Yang-Mills on the cylinder}

In \cite{brownian} a relation was found between Brownian motion/discrete random walk models of $N$ non-intersecting movers on a line, and certain low-dimensional gauge theories. The gauge theories involved were 3d Chern-Simons theory, 2d Yang-Mills, and the plaquette model of QCD$_2$. As it turned out, a fruitful way to think about these
motions is in terms of a single particle performing Brownian motion in the Weyl chamber 
of some (finite or affine) Lie algebra. In this section we will briefly review the basic
correspondence and work out in detail the correspondence between Brownian motion and 2d
Yang-Mills, leaving the Brownian motion description of Chern-Simons theory for a future
publication.

Let us first recall some basic facts about 2d Yang-Mills on the cylinder that we will use later on. The partition function can be shown to be \cite{migdal}
\be\label{2dym}
Z_{\sm{2dYM}}(g,g';t)=\sum_{\l\in P_+}\chi_\l(g^{-1})\chi_\l(g')e^{-\half t\,C_2(\l)}~.
\ee
The Yang-Mills coupling always appears in combination with the area of the cylinder, 
$t=g_{\sm{YM}}^2A$. $C_2(\l)$ is the eigenvalue of second Casimir in the representation $\l$ (for notation and conventions, see appendix \ref{liea}).
The sum runs over all finite dimensional, irreducible representations, labeled by their highest weight.

This partition function is the kernel for the heat equation on the group manifold of the 
corresponding group $G$:
\be
{\pa\over\pa t}\,Z_{\sm{2dYM}}(g,g')=-\half\,C_2(g)Z_{\sm{2dYM}}(g,g')~.
\ee
Given a boundary condition $f_0(g)$ at $t=0$, the heat equation has a unique solution:
\be
f(g,t)=\int\dd g' Z_{\sm{2dYM}}(g,g')\,f_0(g')
\ee
We will later on see that it is important that the coupling constant 
$t=g_{\sm{YM}}^2A$ is complex, and in particular not purely imaginary. The reason is that
the partition function is not analytic when the real part of the coupling goes to zero
\cite{witten}. Thus, one should be careful when continuing solutions of the heat equation
to solutions of the Schr\"odinger equation because in general the continuation is not 
analytic. We will show this explicitly.

The large $N$ limit of 2dYM on the cylinder has been studied in \cite{grossmatytsin}. We
should here also mention the interesting related work \cite{zelditch}.

\subsection{Brownian motion and low-dimensional gauge theories: the basic correspondence}

In \cite{brownian} we considered Brownian motion of a number $r$ non-intersecting movers on a line.
Alternatively, this can be regarded as motion of a single particle in the Weyl chamber of
a Lie group, whose dimension equals the rank of the gauge group, $r$. We will first review the prototypical case of Chern-Simons on $S^3$ with gauge group $U(N)$, in which case $r=N$. In what follows we briefly summarize this relation
\cite{brownian}. Details will be presented elsewhere. 

Even though for some purposes the description in terms of $N$ particles is more useful 
than one where one considers a single mover, in this paper we will describe 
things from
the perspective of a single mover performing Brownian motion in the Weyl chamber of some
Lie algebra. Consider first $U(N)$. Its Weyl chamber is the dual Cartan subalgebra modded out by the
Weyl group, in this case the permutation group $S_N$. We can arrange coordinates so that
the Weyl chamber is given by a vector $\l=(\l_1,\ldots,\l_N)$, where 
$\l_1>\l_2>\ldots\l_N$. The probability for this particle to travel from $\m$ to $\l$ 
in time $t$ is given by \cite{fisher}
\be
p_{t,N}(\l,\m)={1\over(2\pi t)^{N/2}}\,e^{-{|\l|^2+|\m|^2\over2t}}\,
\det|e^{\l_i\m_j/t}|_{1\leq i<j\leq N}~.
\ee
The vectors $\l$ and $\m$ label irreducible representations of $U(N)$ when they lie in
$P_+$, i.e. when their coefficients are positive integers in a basis of fundamental
weights, which we will assume in what follows. Notice, however, that Brownian motion is
a continuous process and so the intermediate steps do not lie on this lattice, but only
inside the Weyl chamber. Now it was
shown in
\cite{brownian} that this is the partition function of Chern-Simons on $S^3$ with gauge
group $U(N)$ when the particle travels around a loop starting and ending at 
$\l=\m=\r$. $\r$ is here the Weyl vector, denoting the trivial representation,
and plays the role of our origin of coordinates on the Lie algebra. By taking
different boundary conditions one was also able to get expectation values of Wilson 
lines,
and precise agreement was found. As noted in \cite{brownian}, the above probabilities 
together with the Boltzmann factor provide a representation of $SL(2,{\mathbb{Z}})$ on 
affine characters and so for boundary conditions corresponding to integrable weights 
they are in 1-1 correspondence with the modular matrices of the WZW model. More 
generally, they are characters of the corresponding gauge group.

An alternative expression for the probability was given:
\be
p_{t,N}(\l,\m)={1\over(2\pi t)^{N/2}}\,e^{-{|\l|^2+|\m|^2\over2t}}\,\sum_{w\in W}
\e(w)e^{(\l,w\m)/t}
\ee
which is obtained by the method of images \cite{fisher}.

\subsection{2d Yang-Mills as a Brownian motion}

In this section we consider a particle moving in the fundamental Weyl chamber of the {\em affine}
Lie algebra of the corresponding group. The structure of this Brownian motion 
is quite interesting, and it turns out to be related to 2d Yang-Mills on the cylinder.

In appendix \ref{liea} we recall some definitions and notation that we use
in what follows.

We will be concerned with Brownian motion in the fundamental Weyl alcove of the
algebra. This is the set of weights, moded out by the affine Weyl group $\ti W$. We recall that the
affine Weyl group also includes translations in the coroot lattice $Q^\vee$:
\be
\ti W=W\ltimes T~,
\ee
where $W$ is the Weyl group. Thus we can characterize the fundamental Weyl 
alcove as  $P/lQ^\vee$, where $l=k+g$ -- here, $k$ is the level and $g$ is the dual 
Coxeter number --, 
moded out by the Weyl group. Notice that the Weyl alcove is actually finite, so we can think of the non-intersecting particles as moving on the compact dual Cartan torus. Therefore, in order to obtain the Brownian motion density
in the Weyl alcove we sum over all the images under this translation \cite{brownian}:
\be\label{qtr}
q_{t,r}(\l,\m)={1\over(2\pi t)^{r/2}}\,\sum_{\g\in lQ^\vee}\sum_{w\in W}
\e(w)e^{-{1\over2t}|\g+\l-w\m|^2}
\ee
where\footnote{Despite the notation, the absolute value bars do not imply taking complex
conjugates, but are simply a Cartesian product in a suitable basis.} 
$|\l|^2=\sum_{i=1}^r\l_i^2$. For $U(N)$, $r=g=N$ and $l=k+N$, but in the following we will not assume this.

By construction, $q_{t,r}(\l,\m)$ still satisfies the heat equation
\be
{\pa\over\pa t}\,q_{t,r}(\l,\m)=\half\D\, q_{t,r}(\l,\m)
\ee
where $\D$ is the flat space Laplacian with respect to the boundary conditions $\l$ or $\m$. 

There is a different way to rewrite the above expression which is in some respects more natural.
It is not hard to see that the following identity holds:
\be\label{qaffine}
\sum_{\g\in lQ^\vee}\sum_{w\in W}\e(w)e^{-{1\over2t}|\g+\l-w\m|^2}
=e^{-{1\over2t}(|\hat\l|^2+|\hat\m|^2)}\sum_{w\in\ti W}\e(w)
e^{{1\over t}(w\hat\l,\hat\m)}
\ee
where we defined affine vectors\footnote{We can include the quadratic factors
in the definition of the affine weights by adding non-zero $n_\l$ pieces. 
However, we prefer to keep these factors explicit. See the next section.} 
$\hat\l=\l+l\hat\o_0$, $\hat \m=\m+l\hat\o_0$. Notice that these vectors 
are in the fundamental alcove of the algebra, as they should. In fact, they are of the 
form $\hat\L+\hat\r$, where $\hat\L$ has level $l$. Thus we get
\be
q_{t,l}(\l,\m)={1\over(2\pi t)^{r/2}}\,\sum_{w\in\ti W}\e(w) 
e^{-{1\over2t}|w\hat\l-\hat\m|^2}~.
\ee
We could have guessed the above expression from the analogy with the 
finite case. In fact, discretized versions of such expressions for the case of random 
walks are mentioned in \cite{gesselzeilberger,grabiner} and analyzed in
\cite{filaseta} in the context of a generalized ballot problem. For Brownian 
motion, \cite{hobson} gives a related expression for motion of $N$ 
non-interacting particles on a circle. The above gives a derivation
based on the compactification of the infinite Weyl chamber.

We now turn to the relation of the Brownian motion density to 2d Yang-Mills 
theory on the cylinder. Our aim is to show how the partition function of 
2dYM, \eq{2dym}, is related to the Brownian motion density, \eq{qaffine}. To
that end we make use of Frenkel's results \cite{frenkel}. We rewrite theorem (4.3.4) of 
\cite{frenkel} as follows:
\be\label{Frenkelsth}
\sum_\n\chi_\n(e^{\l'})\chi_\n(e^{-\m'})e^{-\half t'|\n+\r|^2} =
{(-il)^r\over(2\pi t)^{r/2}}
{|P/Q^\vee|\over D_\r(\l')D_\r(-\m')}\,\sum_{\g\in lQ^\vee} \sum_{w\in W}\e(w)
e^{-{1\over2t}|\g+\l-w\m|^2}
\ee
where $D_\r(\l)$ is the denominator of the character:
\be
D_\r(\l)=\prod_{\a>0}2\sinh(\a,\l)/2~.
\ee
$P$ is the weight lattice, and $|P/Q^\vee|$ is the volume of $P/Q^\vee$, i.e. the number of points of $P$ that are contained within a unit cell of the coroot lattice $Q^\vee$. For the external states on the cylinder we filled in group elements $g'=e^{\l'}$, $g=e^{\m'}$, and we defined
\bea
\l'&=&{2\pi i\over l}\,\l\nn
\m'&=&{2\pi i\over l}\,\m\nn
t'&=&\left({2\pi i\over l}\right)^2t~.
\eea 
The proof, which we will not repeat here, goes in two main
steps. First one uses Weyl's character formula on the left-hand side and reduces the final expression to an expression
involving only a sum over weights and a single sum over elements of the Weyl group. Then one
uses Poisson's resummation formula to replace the sum over the weights by a sum over coroots.

Using this, we find:
\be\label{cylinder}
Z_{\sm{2dYM}}(g,g';t)={(-il)^r|P/Q^\vee|\over D_\r(\l')D_\r(-\m')}
\,q_{t,r}(\l,\m)~.
\ee

This is the expression in \cite{brownian}. Notice that the Poisson resummation effectively transforms the time-dependence as $t\rightarrow-1/t$. 

Notice that the relation between 2d Yang-Mills and Brownian motion involves factors of $D_\r(\l)$ 
that depend on the representation. These come from the normalization of the characters 
$\chi_\l(g)$. The partition function has been normalized so that it has the
expected limit on the disc, i.e. when
$g\rightarrow$ id. Of course, such normalization factors 
are independent of $t$. This normalization ensures that both sides satisfy the heat equation and will be explained in section \ref{verlinde}. On the other hand we see that 
Chern-Simons and Brownian motion share the same normalization.

\subsection{The structure of 2d Yang-Mills on the cylinder}\label{structure}

In \cite{brownian} we claimed that the modular properties of 2d Yang-Mills on
the cylinder are simple to analyze. This follows from the representation
\eq{cylinder} and the analysis is absolutely standard, see e.g. \cite{kac,difrancesco}. Since the 
normalizing prefactor has no $t$-dependence,
we will analyze in detail the expression
\be
q_{t,r}(\l,\m)={1\over(2\pi t)^{r/2}}\,\sum_{\g\in lQ^\vee}\sum_{w\in W}
\e(w)e^{-{1\over2t}|\g+\l-w\m|^2}~.
\ee
This can be expressed in terms of the following theta-function:
\be
\Theta^{(l)}_\l(\zeta;\t;s)=e^{-2\pi ils}\sum_{\a^\vee\in Q^\vee}
e^{\pi il\t|\a^\vee+\l/l -\zeta/\t|^2 -\pi il|\zeta|^2/\t}
\ee
where $l$ is any integer (in our case it will be the level). Taking
\bea\label{modtrafo}
{1\over t}&=&-{2\pi i\over l}\,\t\nn
\m&=&{l\over\t}\,\zeta\nn
s&=&-{|\zeta|^2\over2\t}~,
\eea
we have to evaluate the $\Theta$-function at the point $(\zeta;\t;s)=-{1\over2\pi it}
\left(\m;l;-{|\m|^2\over2l}\right)$. Thus we get the following identity
\be
q_{t,r}(\l,\m)={1\over(2\pi t)^{r/2}}\sum_{w\in W}\e(w)\Theta_{w(\l)}(\zeta;\t;s)~.
\ee
$S$-transformations are generated by $(\zeta;\t;s)\rightarrow\left({\zeta\over\t}
;-{1\over\t};s+{|\zeta|^2\over2\t}\right)$. In 
this case, the $S$-matrix is represented by
\be
S^{(l)}_{\l\m}=\left({-i\t\over l}\right)^{r/2}|P/Q^\vee|^{-1/2}e^{-2\pi i(\m,\l)/l}
\ee
and
\be
\Theta^{(l)}_\l\left({\zeta\over\t};-{1\over\t};s+{|\zeta|^2\over2\t}\right)=\sum_{\m\in P/lQ^\vee}
S^{(l)}_{\l\m}\,\Theta^{(l)}_\m(\zeta;\t;s)~.
\ee
The sum
\be
\f_\l(\zeta;\t;s)=\sum_{w\in W}\e(w)\Theta^{(l)}_{w\l}(\zeta;\t;s)
\ee
transforms under $S$-transformations as:
\be
\f_\l\left({\zeta\over\t};-{1\over\t};s+{|\zeta|^2\over2\t}\right)=\left(-i\t\over l\right)^{r/2}
|P/Q^\vee|^{-1/2}\sum_{w\in W}\e(w)\sum_{\m\in P/lQ^\vee}e^{-2\pi i(w\l,\m)/l}
\Theta_\m^{(l)}(\zeta;\t;s)
\ee
and under $T$-transformations:
\be
\f_\l(\zeta;\t+1;s)=e^{i\pi|\l|^2/l}\,\f_\l(\zeta;\t;s)~.
\ee

It is now straightforward to see how $q_{t,r}(\l,\m)$ transforms.
First of all, let us add the additional parameter $s$ in our notation. This simply keeps track of
the phases generated:
\be
q_{t,r}(\l,\m;s)=e^{-2\pi ils}q_{t,r}(\l,\m)~.
\ee
A modular transformation on $(\zeta;\t)\rightarrow(\zeta';\t')$ redefines $\m,t$ as follows:
\bea
{1\over t'}&=&-\left(2\pi i\over l\right)^2\,t\nn
\m'&=&{\m\over2\pi it}~.
\eea
We get:
\be
q_{t',r}(\l,\m';s')=\left({2\pi t\over l^2}\right)^{r/2} \sum_{w\in W}\e(w)
\sum_{\m\in P/lQ^\vee}S_{w\l,\m}^{(l)}\,\Theta^{(l)}_\m(\zeta;\t;s)~,
\ee
where $S$ is defined as before. This can also be written as
\bea\label{modfinal}
q_{t',r}(\l,\m';s')&=&\left(2\pi t\over l\right)^{r/2}|P/Q^\vee|^{-1/2}
\sum_{\n\in P_{++}^k}\sum_{w\in W}\e(w)\,e^{-{2\pi i\over l}(w\l,\n)}\,q_{t,r}(\n,\m;s)\nn
&=&i^{-|\D|_+}(2\pi t)^{r/2}\sum_{\n\in P_{++}^k} S_{\l\n}\,q_{t,r}(\n,\m;s)
\eea
where $S$ is now the usual, unitary representation of the $S$-matrix on affine characters. Notice that in our case we start with the particular value of $s$ 
\eq{modtrafo} and end up with
$s'=0$. Now because of \eq{cylinder}, this gives us the transformation rules
of $Z_{\sm{2dYM}}(g,g';t)$. We see that $q_{t,r}$ transforms in a unitary representation
of $SL(2,{\mathbb{Z}})$, up to a weight of $r/2$.

We get a unitary representation of the modular transformations if we appropriately normalize the
partition function. Indeed, it should by now be clear that
\be
{q_{t,r}(\l,\m)\over q_{t,r}(\r,\m)}= e^{-{1\over2t}C(\l)}\,
\chi_{\hat\l}(\hat\m/t)
\ee
where the affine character is
\be
\chi_{\hat\l}(\m)={{\sum_{w\in\hat W}}\e(w)e^{(w\hat\l,\hat\m)}
\over \sum_{w\in\hat W}\e(w)e^{(w\hat\r,\hat\m)}}~.
\ee
$\hat\l$ and $\hat\m$ are defined as in \eq{qaffine}. In our case the modular
anomaly term is absent because\footnote{It will give a non-vanishing contribution, however, if we decide to include the Casimir in the definition of the affine weights, as discussed earlier.} $n_\l=n_\m=0$. The above formula is completely analogous to Brownian motion in the Weyl
chamber of the classical group, for example $U(N)$, where we found that the normalized probability 
was the character of $U(N)$. The modular transformation properties of the affine character are 
now the usual ones:
\bea\label{modularST}
\chi_{\hat\l}(\zeta;\t+1;s)&=&\sum_{\hat\m\in P_+^k}T_{\hat\l\hat\m}\,\chi_{\hat\m}(\zeta;\t;s)\nn
\chi_{\hat\l}\left({\zeta\over\t};-{1\over\t};s+{|\zeta|^2\over2\t}\right)&=&\sum_{\hat\m\in P_+^k}
S_{\hat\l\hat\m}\,\chi_{\hat\m}(\zeta;\t\s)~,
\eea
where $S$ and $T$ are the usual representations of the modular matrices on affine characters, given in appendix \ref{liea}.

\section{Two-dimensional Yang-Mills on the cylinder and Chern-Simons on 
$S^3/{\mathbb{Z}}_q$}

Some experience with the modular matrices of the WZW model and $U(N)$ characters immediately tells
that, up to a normalization, an object such as \eq{2dym} is very similar to a product of 
modular matrices $ST^qS$ for some power $q$ related to the YM coupling. There are, however, important differences. The first one is that
the Hilbert spaces are different. In 2d Yang-Mills, we sum over all finite dimensional
representations, and also the external states at the two ends of the cylinder can be in arbitrary
representations. The WZW model, on the other hand, is based on an affine Lie algebra where the
Hilbert space only contains integrable representations satisfying 
$(\l,\th)\leq l$, where $\th$ is the highest root. This Hilbert space is finite.
The second difference is in the coupling
constant. 2d Yang-Mills really satisfies a heat equation, whereas $ST^qS$, if $q$ or the level
were continuous parameters, would satisfy the Schr\"odinger equation. Nevertheless one would
expect that some kind of analytic continuation should be possible, in the same way that we can
describe topological strings by identifying the string coupling as $g_s={2\pi i\over k+N}$.

Of course, we expect that in the large $k$ limit, and up to possible divergences,
2dYM and Chern-Simons agree, as in that case the Hilbert spaces are the same. This is 
indeed the result of \cite{witten2d,jones,blauthompson}.

The purpose of this section is to make the above observations on the finite $k$ case 
precise. To this end, it will prove
far more useful to start from the expression \eq{cylinder} than to work directly with \eq{2dym}.
Basically, in \eq{cylinder} the two characters have already been expanded into a single expression of the
type found in the Chern-Simons literature.

It turns out that there are
special values of the coupling $g_{\tn{YM}}$ at which the partition function \eq{2dym} collapses
to the gluing operator $ST^qS$ in the way expected, i.e., up to normalization. These values of the
coupling are imaginary, in particular they are $2\pi i$ times a rational number, and precisely at these values the sum over all representations
reproduces the sum over the integral representations in Chern-Simons. This uses a certain
periodicity of the partition function for these values of the coupling. Since the partition function is periodic on an infinite lattice, there is an overall diverging factor
multiplying the partition function that we will explain.

\subsection{Chern-Simons on $S^3/{\mathbb{Z}}_q$}

In \cite{jones}, a surgery prescription was constructed for obtaining partition functions and expectation values of Wilson loops on non-trivial manifolds. One glues together a three-manifold ${\cal M}$ out of 
two solid tori by appropriate identification of the $A$- and $B$-periods of the two tori after diffeomorphisms.
These diffeomorphisms are represented by $SL(2,{\mathbb{Z}})$ transformations. 
The partition function of Chern-Simons theory on this manifold is then given by the representation
of the corresponding gluing operator on the Hilbert space of the WZW model.

Chern-Simons theory in lens spaces was introduced in \cite{jones} and 
studied in \cite{jeffrey,rozansky,marcos,akmv}. The corresponding matrix model 
for the case $S^3/{\mathbb{Z}}_q$ was analyzed in \cite{akmv,nick}. It was shown in
\cite{akmv} that modding out the $S^3$ by ${\mathbb{Z}}_q$ corresponds to
acting on the cycles of the torus before gluing by an $SL(2,{\mathbb{Z}})$
matrix $U=ST^qS$. If we represent a generic $SL(2,{\mathbb{Z}})$ matrix by
\be
U^{(p,q)}=\left(
\begin{array}{cc}
p & r \\
q & s
\end{array}
\right)~,
\ee
and in the usual representation
\bea
T&=&\left(
\begin{array}{cc}
1 & 1 \\
0 & 1
\end{array}
\right)~,\nn
S&=&\left(
\begin{array}{cc}
0 & -1 \\
1 & \,\,\,0
\end{array}
\right)~,
\eea
then it follows that we need\footnote{In this, our conventions differ
from those in \cite{akmv}, where all signs in $U$ were positive. Of course this is a trivial redefinition, as one can explicitly show. In fact one matrix is minus the inverse of the other.} $U^{(-1,q)}=ST^qS$ with $p=s=-1$, $r=0$. Flipping the signs of $p$ and $s$ would simply
correspond to flipping the sign of $g_{\tn{YM}}$ and of the external states. We will not consider this in this paper. 

The representation of the generic operator $U^{(p,q)}$ in this Hilbert space
is given in \cite{jeffrey,rozansky,marcos}. In the case of interest to us we have:
\be\label{gluing}
U_{\l\m}^{(-1,q)}={[i\,\mbox{sgn}(q)]^{|\D_+|}\over(l|q|)^{r/2}} 
{e^{-{\pi i d\over12}\,\F(U^{(-1,q)})} \over|P/Q^\vee|^{1/2}} \sum_{\g\in Q^\vee/qQ^\vee}
\sum_{w\in W}\e(w)e^{-{\pi i\over lq}|l\g+\l+w\m|^2}~.
\ee
Here, $\F(U^{(p,q)})$ is the Rademacher function, related to a choice of 
framing \cite{atiyah,witten,jeffrey}. In this paper we will not keep track of 
constant phase factors, although this would be an important check.

As shown in \cite{rozansky}, taking matrix elements of $U^{(-1,q)}$ in the 
trivial representation gives the partition function of Chern-Simons on 
$S^3/{\mathbb{Z}}_q$. Taking it between non-trivial representations $\l$ and
$\m$ means that we glue two tori with colored cycles $\l$ and $\m$, which results in the expectation value of a link $W_{\l\m}$ embedded in $S^3/{\mathbb{Z}}_q$.

As a trivial check, the geometric picture tells us that the case $q=1$ should give us 
back the three-sphere. This is indeed the case, as we can explicitly see from 
\eq{gluing}, which (up to framing) reduces to $T^{-1}ST^{-1}$. This is nothing else than
$STS$, as it should.

We can already see the very similar structure with \eq{qtr}. In fact, under 
the identification 
$1/t=2\pi i/lq$, up to numerical factors independent of $l$ and $q$ the only difference 
seems to 
be the summation range. One would think that in the limit $q\rightarrow\infty$ both expressions 
precisely agree. Things are nevertheless a bit more subtle. The expression \eq{qtr} does not 
make sense when $1/t=2\pi i/lq$. Due to the periodicity of the lattice, an infinite number of 
images will contribute exactly the same weight. That is why, whereas the real 
expressions are finite, continuation to complex values of the coupling is 
subtle in this case. In fact we will exploit this periodicity to show 
that, up to this divergence, both expressions do agree without the need to take $q$ to infinity. The divergence is well-known in 2dYM. The partition function of 2dYM is in 
general non-analytic as $g_{\tn{YM}}\rightarrow0$ (depending on the Riemann surface and
gauge group) \cite{witten}. We seem to find here that it stays non-analytic for these
special values of the imaginary part of the coupling, whenever 
${\mbox{Re}}\,g_{\tn{YM}}=0$. Although from the point of view of 2dYM it is not always
 natural to remove these divergences \cite{witten2d,witten}, they can be consistently 
absorbed in the definition of the zero-point energy, as we will discuss later. 

From the Brownian motion point of view, we might say that complexifying time as above 
effectively amounts to 
putting periodic boundary conditions inside the Weyl chamber, so that we are restricting
to the integral weights. All the paths outside this domain then contribute with the same
weight.

\subsection{Relation between 2d Yang-Mills and Chern-Simons}

Let us now see how we can continue the partition function to imaginary values
of the coupling. This has to be done carefully, as it diverges on the imaginary axis. We
can rewrite the partition function as
\bea\label{partf}
Z_{\sm{2dYM}}(g,g';t')&=&\sum_\n\chi_\n(e^{\l'})
\chi_\n(e^{-\m'}) e^{-\half t'(C_2(\n)+|\r|^2)}\nn
&=&\left({-l^2\over2\pi t}\right)^{r/2}{|P/Q^\vee|\over D_\r(\l')D_\r(-\m')}
\sum_{\g\in lQ^\vee}\sum_{w\in W}\e(w) 
e^{-{1\over2t}|\g+\l-w\m|^2}\nn
&=&\left({-l^2\over2\pi t}\right)^{r/2}{|P/Q^\vee|\over D_\r(\l')
D_\r(-\m')}\sum_{w\in W}\e(w) e^{-{1\over2t}|\l-w\m|^2}\,z(\l-w\m)
\eea
where for convenience we included an extra factor of $|\r|^2$ in the exponent,
and as before $\l'=2\pi i\l/l$, $\m'=2\pi i\m/l$, and $t'=(2\pi i/l)^2t$. We 
also defined the function
\be
z(\l)=\sum_{\g\in lQ^\vee}e^{-{1\over2t}(\g^2+2\g\l)}~.
\ee
This is a higher-dimensional generalization of the usual theta function, whose 
asymptotic properties are well-known. It is well-known to converge when $i/t$ 
is in the upper-half plane, and for any $\l\in{\mathbb{C}}$.

We will be interested in the asymptotic properties of this function. To get the partition function 
from this we just multiply by the relevant Gaussians and sum over images. Since we want to compare 
with Chern-Simons theory, we need to identify
\be
{1\over t}={2\pi i\over lq}+{2\e\over l}
\ee
where $\e>0$ is a real constant that we will take to zero at the end of the
computation. We can rewrite the above as
\be
z(\l)=\sum_{\g\in Q^\vee}e^{-{i\pi\over q}(l\g^2+2\g\l)-\e(l\g^2+2\g\l)}~.
\ee
We now split the sum as follows: write $\g=\a^\vee+q\b^\vee$ where
$\a^\vee\in Q^\vee/qQ^\vee$, $\b^\vee\in Q^\vee$, so
\bea
z(\l)&=&\sum_{\a^\vee\in Q^\vee/qQ^\vee}\sum_{\b^\vee\in Q^\vee}\exp[-i\pi(l
\a^\vee{}^2/q+lq\b^\vee{}^2+2l\a^\vee\b^\vee +2\a^\vee\l/q+2\b^\vee\l)+\nn
&&-\e(l\a^\vee{}^2+2\a^\vee\l+q^2\b^\vee{}^2+2q\b^\vee (\a^\vee+\l))]~.
\eea
Now the following conditions hold\footnote{Notice that \cite{frenkel} defines the 
weights with a factor of $2\pi i$, which explains our normalization $\l'=2\pi i\l/l$.
The rescaling by $l$ was chosen for convenience.}:
\bea
\g^2&\in&2{\mathbb Z}_+\nn
\g\l&\in& l{\mathbb Z}~.
\eea
We can use this property to reduce the sums. The only terms that remain are:
\be\label{z}
z(\l)=\sum_{\a^\vee\in Q/qQ^\vee}\sum_{\b^\vee\in
Q^\vee}e^{-\e[q^2\b^\vee{}^2
+2q\b^\vee(\a^\vee+\l)]-{i\pi\over q}(l\a^\vee{}^2+2\a^\vee\l)
-\e(l\a^\vee{}^2+2\a^\vee\l)}~.
\ee
Obviously, the last term proportional to $\e$ we can drop, as it has no
$\b^\vee$-dependence and only gives a
finite contribution that will go to zero when we take $\e\rightarrow0$.
The second term in the exponential is the one we are interested in; it is 
the one related to $U$. The first term is there only to 
regularize the sum and contributes an infinite normalization constant. Clearly, the diverging
contribution comes from the infinite sum over $Q^\vee$ in the limit $\e\rightarrow0$. 
The 
contribution of this term is analyzed in the appendix \ref{thetaf}. Using those results, we
finally get:
\be\label{zfinal}
z(\l)=\left({\pi\over q^2\e}\right)^{r/2}{1\over|P/Q^\vee|^{1/2}}\,
\sum_{\a^\vee\in Q^\vee/qQ^\vee}e^{-{i\pi\over q}(l\a^{\vee2}+2\a^\vee\l)}~.
\ee
Filling this into the summation in \eq{partf}, we get for the leading divergence
\be
\sum_{\g\in lQ^\vee}\sum_{w\in W}\e(w)e^{-{1\over2t}|\g+\l-w\m|^2}
=\left({\pi\over q^2\e}\right)^{r/2}{1\over|P/Q^\vee|^{1/2}}
\sum_{\g\in Q^\vee/qQ^\vee}\sum_{w\in W}\e(w)e^{-{i\pi\over lq}|l\g+\l-w\m|^2}~.
\ee
The last term is the gluing operator \eq{gluing}, where the irrelevant sign 
difference in $\m$ comes from the fact that at one end of the cylinder we are 
considering the conjugate representation $g^{-1}=e^{-\m'}$.

Filling this into the partition function \eq{partf}, we get
\be
Z_{\sm{2dYM}}(g,g';t)=\left({-i\pi l\over q^3\e}\right)^{r/2}
{|P/Q^\vee|^{1/2}\over D_\r(\l')D_\r(-\m')} \sum_{\g\in Q^\vee/qQ^\vee}
\sum_{w\in W}\e(w)e^{-{i\pi\over lq}|l\g+\l-w\m|^2}~.
\ee
This is now, up to normalizations, the gluing operator $U^{-1,q}_{\l\m}$. As mentioned at the
beginning of this section, the normalization by factors of $D_\r$ has to do with the normalization
of the character in 2d Yang-Mills. This is different from the normalization of $S$, which does not
contain that factor in the Weyl denominator formula. If we normalize the gluing operator the same 
way as the characters, that is, such that the $S$-matrices are normalized to unit strenght, we get
\be\label{ulm}
u_{\l\m}={(ST^qS)_{\l\m}\over S_{\l\r}S_{\r\m}} =(l/|q|)^{r/2}{(-i\,{\mbox{sgn}}(q))^{|\D_+|}\over D_\r(\l')D_\r(-\m')}|P/Q^\vee|^{1/2} \sum_{\g\in Q^\vee/qQ^\vee}
\sum_{w\in W}\e(w)e^{-{i\pi\over lq}|l\g+\l-w\m|^2}~.
\ee
Thus, we finally obtain:
\be\label{ulm2}
u_{\l\m}=e^{i\f}\left(q^2\e\over\pi\right)^{r/2}\, 
Z_{\sm{2dYM}}(e^{\l'},e^{\m'}; t')
\ee
where $\f$ is a phase that we are not considering in this paper\footnote{This
phase is related to the choice of framing in Chern-Simons and would be 
interesting to compute.}. The quantity on the right-hand side is finite. The fact that 
the normalizations now agree precisely matches our naive expectations.

Coming back to the case of $S^3$, it is now clear that the transformation
$t\rightarrow-1/t$ in \eq{Frenkelsth} means that $STS=T^{-1}ST^{-1}$, as we see from the 
above.

The divergence is not new and can now be absorbed as a constant in the definition of the zero-point energy. 
Notice that it depends only on $q$ and $\e$ but not on any other quantities. Recalling that the
action was
\be
S_{\sm{2dYM}}=-{1\over2g_{\tn{2dYM}}^2}\int_\S\dd^2x\sqrt{g}\,\Tr F^2~,
\ee
we can add a counterterm given by
\be\label{ct}
S_{\sm{ct}}=c\int_\S\dd^2x\sqrt{g}
\ee
and we see that we need to adjust $c={r\over2A}\log{q^2\e\over\pi} =lqr{g_{\tn{YM}}^2
\over4\pi i}\log{q^2\e\over\pi}$. Such counterterms are
invariant under area-preserving diffeomorphisms and hence respect the symmetries of the 
theory. In our case, the area drops out in \eq{ct}, and the counterterm is really a pure number, because
$\e$ is. Both $g_{\tn{YM}}$ and $A$ are dimensionful, but their product is 
dimensionless:
\be
g_{\tn{YM}}^2A={2\pi i\over lq}-{2\e\over l}~.
\ee
Thus, although we have fixed the product, we still have the freedom to rescale 
$g_{\tn{YM}}^2\rightarrow g_{\tn{YM}}^2t$, $A\rightarrow A/t$, thus preserving the 
symmetry.

Counterterms that redefine the zero point energy and thus shift the partition 
function by a constant have been discussed in \cite{witten2d,witten}. There, 
it was discussed that there is an ambiguity in the definition of the partition 
function related to the addition of such terms\footnote{On general Riemann 
surfaces, there is a second counterterm involving the Euler characteristic.}, 
different regularization schemes giving different values of $c$. This divergence is
however just a manifestation of the non-analytic structure of 2dYM.

Notice that the divergent prefactor of $(\e/\pi)^{r/2}$ in \eq{ulm2} is precisely the factor of 
$(2\pi t/l)^{r/2}$ that we got in our final expression for our modular 
transformation 
\eq{modfinal} if we subtract the imaginary part of $t$. Also,
notice that shifts of $\t\rightarrow\t+1$ correspond to shifts of $t$ by $2\pi i/l$. This suggests that the factor that we found by explicit computation might
be related to the prefactor that follows from the modular transformation 
properties, although we have not checked this explicitly. In particular, the
appropriate $q$-dependence should be taken into account. It might be possible
to use the modular transformations in section \ref{structure} to get the 
above results directly. 

Let us comment on the normalization of the operator $u_{\l\m}$ that we used in \eq{ulm}.
We will see that this normalization agrees with the results in \cite{witten2d,witten}. 
Whereas
the normalization in 2dYM is different from the normalization of $U^{(-1,q)}$, Brownian motion
actually automatically comes with the same normalization as $U^{(-1,q)}$. This was expected too, as it was found in \cite{brownian} that, up to framing
factors, these normalizations generically agree in the case of Brownian motion in the Weyl chamber of the finite algebra, $S$ having the 
interpretation of Brownian motion densities. In particular the
above implies that one should 
be able to obtain $q_{t,r}(\l,\m)$ from a composition of two finite probabilities $p_{t,r}$ with a 
Boltzmann factor, and summing over intermediate states. This should be most clear in the matrix 
model representation.

\section{2dYM, the WZW model, and the Verlinde formula}\label{verlinde}

We will now comment on the results in the previous section from a different point of view, mentioned in the introduction. In a beautiful series of papers, Witten \cite{witten2d,witten} and Blau and Thompson \cite{blauthompson} have related 2dYM/BF theory to Chern-Simons theory and G/G WZW models. An important upshot of this was in using the Verlinde formula \cite{erik} in combination with the Riemann-Roch formula to recover the volumes of moduli spaces of flat connections on Riemann surfaces:
\be\label{riemannroch}
{\mbox{dim}}\,{\cal M}(\S_g)=\lim_{k\rightarrow\infty}k^{-\half{\sm{dim}}\,
{\cal M}(\S_g)}\,{\mbox{dim}}\,V_{g,k}~.
\ee
The left-hand side was given by the partition function of 2dYM in the topological limit or, alternatively, by topological BF theory. Let us recall the definition:
\bea
Z_{\sm{2dYM}}(\S_g,g_{\sm{YM}})
&=&\int{\cal D}A \exp\left(-{1\over8\pi^2g^2_{\tn{YM}}} \int_{\S_g}\Tr F*F\right)\nn
&=&\int{\cal D}A{\cal D}\f \exp\left({i\over4\pi^2}\int_{\S_g}\Tr\f F
-{g^2_{\tn{YM}}\over8\pi^2}\int_{\S_g}\dd\m\,\Tr\f^2\right)
\eea
In the last line we rewrote 2dYM in terms of BF-theory with a quadratic potential. Now the point is that in the limit $g_{\tn{YM}}\rightarrow0$ the theory is topological, and the path integral reduces to a delta function with support on flat connections. Since we integrate over all gauge connections, the result is the left-hand side of \eq{riemannroch}. Now the right-hand side is given in conformal field theory by the modular $S$-matrix:
\be\label{dim}
{\mbox{dim}}\,V_{g,k}=\sum_\l S_{\r\l}^{2-2g}~.
\ee
It is after all not surprising that this is related to 2dYM if we consider the well-known fact that the quantum dimensions reduce to the dimensions of representations of the gauge group in the large $k$ limit:
\be\label{diml}
{\mbox{dim}}\,\l=\lim_{k\rightarrow\infty}{S_{\l\r}\over S_{\r\r}},
\ee
and the partition function of 2dYM on a Riemann surface of genus $g$ is, in the limit $g_{\tn{YM}}\rightarrow0$:
\be
Z_{\sm{2dYM}}(\S_g,0)=\sum_\l({\mbox{dim}}\,\l)^{2-2g}~.
\ee
This equation is however only formal, because this limit is in general not well-defined. As we already mentioned, in certain cases 2dYM is non-analytic in the coupling constant, and this is an indication of the singular nature of the moduli space \cite{blauthompson}.

If we include $s$ punctures with holonomies around them and label them by representations $\m_i$, $i=1,\ldots,s$, the formula for the Verlinde numbers becomes:
\be\label{V}
{\mbox{dim}}\,V_{g,s,k}=\sum_\l S_{\l\r}^{2-2g-s}\prod_{i=1}^sS_{\l\m_i}~.
\ee

There is a generalization of the above to finite $k$ due to Blau and Thompson \cite{blauthompson}. They show that the G/G WZW model on $\S_g$, which is equivalent to Chern-Simons on $\S_g\times S^1$, is equivalent to BF theory with a compact scalar. Roughly speaking, when we take $k\rightarrow\infty$ the scalar becomes non-compact and we get the topological YM/BF theory back. Obviously, at finite $k$ the partition function of this compact BF theory is no longer a delta-function. Nevertheless this still computes \eq{dim}, i.e. the dimensions of the spaces of conformal blocks of the WZW model when $k\rightarrow\infty$.

In the case of the cylinder we have been considering in this paper, the formula \eq{V} is rather trivial: it reduces to $(S^2)_{\m_1\m_2}=\d_{\m_1\m_2}$, which is simply the satement that the Hilbert space of Chern-Simons theory on $\S\times{\mathbb{R}}$ (the Hilbert space of conformal blocks on $\S$) when $\S$ is a cylinder is one-dimensional if $\m_1$ and $\m_2$ are conjugate representations, and zero otherwise \cite{jones}.

As we have seen, however, we do {\it not} need to consider $g_{\tn{YM}}$ to be small to get back quantities that have an interpretation in the WZW model at finite $k$. This can be seen as a generalization of the relation in \cite{witten2d,blauthompson} between 2dYM and Chern-Simons/WZW to finite $k$. Indeed, we have seen that the partition function of 2dYM computes matrix elements of the type $ST^qS$. Notice that the issue of normalization was exactly the same we encountered above for the $k\rightarrow\infty$ case: 2dYM quantities naturally come normalized as in \eq{diml}, which are well-defined at large $k$, whereas Chern-Simons naturally computes matrix elements of the type \eq{dim}, which in particular diverge in the classical limit. That is where the normalization \eq{riemannroch} comes from. The situation for 2dYM with coupling $g_{\tn{YM}}^2A={2\pi i\over (k+g)q}$ instead of zero is exactly the same: we need to normalize a matrix element $(ST^qS)_{\l\m}/S_{\l\r}S_{\m\r}$ to be able to compare the theories.

This seems to suggest that the limit of 2dYM where the imaginary part of the coupling is (up to a factor of $2\pi$) a rational number, and the real part goes to zero, is still some sort of topological limit of the theory, precisely analogous to the $g_{\tn{YM}}\rightarrow0$ limit, but not a {\em classical} limit. In fact also in this case we can write down a BF theory
\be
S_{\sm{BF}}={i\over4\pi^2}\int_{\S_g}\Tr\f F -{g^2_{\tn{YM}} \over8\pi^2} \int_{\S_g}\dd\m\,\Tr\f^2
\ee
corresponding to this YM theory. Notice that, in contrast to \cite{blauthompson}, the scalar is now non-compact.

Let us first look at Chern-Simons theory on the sphere, $S^3$. This case is rather trivial; as mentioned before, it corresponds to 2dYM on the cylinder with trivial representations at the two ends and coupling $g_{\tn{YM}}^2A={2\pi i\over k+g}$. In \cite{akmv} it was found by mirror symmetry that it is described by BF theory with a compact scalar and quadratic potential. Notice however that this theory lived on a ${\mathbb{P}}^1$. This is a natural generalization of the compact BF theory found in \cite{blauthompson}, which lived on $\S$ and described Chern-Simons on $\S\times S^1$, to the case of Chern-Simons on $S^3$: one adds a quadratic potential, which corresponds to the gluing operator $S$. In our case, the Yang-Mills theory lives on the cylinder, whereas the scalar is non-compact, and the potential is quadratic as well. Notice that the gauge coupling is the same as in \cite{akmv}. It would be interesting to show that these theories are the same at the level of the path integral. Notice that our cylinder seems to be the same as the cylinder on the B-model side in \cite{akmv}. On the matrix model side, although the measure was unitary, the branes did not go back to themselves after a shift of $2\pi$ around the cylinder.

In the other cases $S^3/{\mathbb{Z}}_q$, the potential of the BF theory that we obtain is the same but divided by $1/q$. This is completely analogous to the situation in \cite{akmv}, where the matrix model potential had a factor of $1/q$.

\section{Discussion and outlook}

Two-dimensional Yang-Mills on a manifold of non-trivial topology is a simple
but still interesting gauge theory. Besides the interest that it has on its 
own and its traditional string interpretation \cite{gross,grosstaylor}, recent 
developments mentioned in the introduction 
\cite{vafa,brownian,nekrasov,mmo,robbert} indicate that it also has useful
applications in 4d SYM and topological strings. According to \cite{vafa}, it
may even tell us something about black holes.

The topological A-model is effectively described by Chern-Simons theory at 
large $N$. If,
on the other hand, topological amplitudes on certain local threefolds can be described by 2dYM \cite{vafa}, one would expect to find direct
relations between Chern-Simons theory and 2dYM. It was
already known that topological 2dYM describes the large $k$ limit of Chern-Simons on 
manifolds of the type $\S\times S^1$ and $\S\times{\mathbb{R}}$. In this paper we have
shown that there is another sector of 2dYM on the cylinder -- that where the coupling is 
$2\pi i$ times a rational number -- which describes Chern-Simons on $S^3/{\mathbb{Z}}_q$ 
at finite $k$, $N$. This Yang-Mills theory can be rewritten as a usual BF theory with a
quadratic potential, and it would be interesting to investigate this theory.

The $S$- and $T$-transformations that we found might also shed some light on the properties of the theory at these special values of the coupling. It would be interesting to see
whether the imaginary part of $g_{\tn{YM}}$ can be interpreted
as a $\th$-angle in YM/BF theory.

Although we did not take the 't Hooft limit for the coupling, this can still naturally
be done in our framework as long as $k$ is not much larger than $N$, since 
$g_{\tn{YM}}^2$ was proportional to ${1\over l}={1\over k+N}$. This gives some hope that 
the matrix model of 2dYM might have some relevance to Chern-Simons in lens spaces.
This matrix model will however have different properties from the traditional 2dYM
matrix model, which generically has a phase transition at finite values of the {\em real}
part of the 't Hooft coupling\footnote{One should however keep in mind the
results in \cite{zelditch}.}. In our case, this coupling is purely imaginary. Besides,
one will have to take into account the divergences. Such
matrix model might nevertheless still be useful, as the natural matrix model that one finds on the 
Chern-Simons side becomes
increasingly complicated for larger values of $q$ \cite{akmv,nick}. Notice
that our relation $g_{\tn{YM}}^2A={2\pi i\over lq}$ precisely matches the
expectation for the topological string, where the effective string coupling is
$g_s={2\pi i\over lq}$ 
\cite{akmv}. It would be extremely interesting to see whether the present
relation between Chern-Simons and 2dYM can be obtained geometrically on the
topological strings side. We already saw that the compact BF theory is obtained by
considering branes in the B-model side \cite{akmv}, and it would be interesting to show
that both theories are equivalent.

We saw that the Hilbert spaces of this 2dYM and of the WZW model basically agree. The 
periodicity of the exponents with respect to shifts of the coupling by these imaginary 
values, and the fact that we are considering special external states at the two ends of 
the cylinder, precisely conspire so that the infinite sum on the YM side
collapses to a sum over the finite Hilbert-space of Chern-Simons theory and 
get the WZW matrix elements.

An interesting extension of our results would be to include the $\th$-angle
in the $U(N)$ case as in \cite{vafa}. This would presumably result in 
additional linear terms in Frenkel's theorem \cite{frenkel}, and the 
generalization should be more or less straightforward. This $\th$-angle plays a
crucial role in the relation with topological strings.

Another interesting extension of this work would be to consider other Riemann surfaces.
At least naively, we do not really expect the present computations to go through for a 
Riemann surface of
arbitrary genus and number of punctures, because adding handles or punctures adds factors
of ${\mbox{dim}}\,\l$, whereas in matrix elements involving products of $S$ and $T$ we
expect the quantum dimensions. However, we do expect that these ideas can be generalized
if we consider products of characters. The Boltzmann factors should then determine the
value of the Yang-Mills coupling.

In \cite{jeffrey}, explicit expressions have also been found for Chern-Simons
on torus fibrations. The expression for the partition function does not involve 
any external states and it bears some resemblance with 2dYM on the torus. It
would be interesting to see whether these expressions can be related to 2dYM.

Finally, let us mention that the Brownian motion description of 2dYM made its
modular transformation properties and some of its analytic structure absolutely explicit 
and provided us with some tools that were useful in relating 2dYM to Chern-Simons theory. In fact we could
say that Brownian motion densities compute correlators in the WZW model. In \cite{dst} it
has been suggested that Brownian motion could be used to compute certain correlators
that involve two-matrix models. It would be interesting to explore these connections further.

\section*{Acknowledgments}

It is a pleasure to thank Stefan Theisen and Miguel Tierz for valuable comments
and collaboration at various stages of this work, and David Grabiner and
Ira Gessel for explaining to us aspects of their work. We would also like to
thank Mina Aganagic, Jos\'{e} M.~Isidro, Marcos Mari\~no, So Matsuura, Kazutoschi Ohta and Annamaria Sinkovics for discussions.

\appendix

\section{Lie algebra conventions}\label{liea}

In this appendix we give some of the definitions and conventions used in the
paper.

In this paper we adhere to the nomenclature that calls the affine Weyl chamber
a ``Weyl alcove'', and reserves the name of ``chamber'' for the finite case.

The Weyl alcove $\hat C_w$ is defined analogously to the Weyl chamber of the finite group $G$:
\be
\hat C_w=\{\hat\l|(w\hat\l,\a_i)\geq0,i=0,\ldots,l\},\,w\in\hat W
\ee
where $\ti W$ is the Weyl group, and $\hat\l$ is an affine weight.
The fundamental Weyl alcove is given by taking $w=1$. Weights in the fundamental alcove are of
the form
\be
\hat\l=\l+k\hat\o_0+r\d~,
\ee
where $k$ is the level, and with $\l\in C_0$ (the fundamental Weyl chamber), 
$r\in{\mathbb R}$ and $\d=(0;0;1)$, $\hat\o_0=(0;1;0)$. $P$ is the weight
lattice, $P_+$ the set of dominant (or highest) weights, and $P_{++}$ are the
regular weights, i.e. they are related to those in $P_+$ by $\l=\m+\r$ where
$\l\in P_{++}$, $\m\in P_+$. $P_+^k$ are the integrable highest weights, so
$(\l,\th)\leq k$.

For completeness, we give the explicit expressions for the $S$- and $T$-modular matrices used in \eq{modularST}. They are represented on
affine characters as follows:
\bea
T_{\lambda \mu } &=&\delta _{\lambda \mu }\,e^{{\frac{2\pi iC(\lambda )}{%
2(k+g)}}-{\frac{2\pi ic}{24}}}\nn
S_{\lambda \mu } &=&{\frac{i^{|\Delta _{+}|}}{(k+g)^{r/2}}}\,|P/Q^{\vee 
}|^{-{\frac{1}{2}}\,}
\sum_{w\in W}\epsilon (w)e^{-{2\pi i\over k+g}(\lambda ,w\cdot \mu )}~,
\eea
where the central charge is $c=k\,{\mbox{dim}}\,g/(k+g)$.

\section{Divergence of the partition sum}\label{thetaf}

In this appendix we analyze the leading divergence in \eq{z}:
\be\label{z2}
z(\l)=\sum_{\a^\vee\in Q/qQ^\vee}\sum_{\b^\vee\in
Q^\vee}e^{-\e[q^2\b^\vee{}^2
+2q\b^\vee(\a^\vee+\l)]-{i\pi\over q}(l\a^\vee{}^2+2\a^\vee\l)}~,
\ee
where we already dropped a contribution that vanishes in the limit $\e\rightarrow0$.
The divergence in this expression comes from the
infinite sum over $Q^\vee$ in the limit $\e\rightarrow0$. In this limit, \eq{z2} is 
completely independependent of $\a^\vee$ and so the expression is not well-defined.

Concentrating on the 
diverging sum over
$Q^\vee$, the relevant $\e$-dependent contribution will be
\be
f(\e,\l)=\sum_{\b^\vee\in Q^\vee}e^{-\e[q^2\b^\vee{}^2+2q\b^\vee\l]}~,
\ee
where we shifted $\l\rightarrow\l-\a^\vee$. 

The above can be written as
\be
f(\e,z)=\sum_{n\in{\mathbb Z}^r}e^{-\e[q^2nCn+2qnz]}
\ee
where $z$ is linear in the $\l$'s and $C$ is the Cartan matrix. This is a generalized 
theta function\footnote{We actually use the notation in \cite{elizalde}.} 
\cite{mumford}:
\be
\Theta\left[{a\atop b}\right](0|\t)_\O =\sum_{n\in{\mathbb Z}^r}\exp\left[-\pi \t(n+a)
\cdot\O\cdot(n+a) +2\pi in\cdot b\right]
\ee
It satisfies:
\be
\Theta\left[{a \atop b}\right](0,\t)_\O ={e^{-2\pi iab}\over \t^{r/2}\sqrt{\det\O}}\,
\Theta\left[{b \atop -a}\right](0,1/\t)_{\O^{-1}}~.
\ee
So we have
\bea
f(\e,z)&=&\Theta\left[{0\atop -\e qz/\pi i}\right](0,\e q^2/\pi)_C\nn
&=&{1\over(\e q^2/\pi)^{r/2}\sqrt{\det C}}\, \Theta\left[{-\e qz/\pi i\atop 0}\right]
(0,\pi/\e q^2)_{C^{-1}}~,
\eea
and we used the fact that the Cartan matrix is non-degenerate. Notice that in the limit 
$\e\rightarrow0$ only the $m=0$ term contributes to the $\Theta$ function, hence up to 
finite terms:
\be
f(\e,z)=\left({\pi\over q^2\e}\right)^{r/2}{1\over|P/Q^\vee|^{1/2}}~,
\ee
so actually the final expression is independent of $z$. This is the result we used in \eq{zfinal}.

\end{document}